\documentclass[paper=a4, fontsize=12pt]{scrartcl}

\usepackage{amssymb}
\usepackage[latin1]{inputenc}
\usepackage[pdftex]{hyperref}

\newcommand{\T}{{\rm Tr}}

\newcommand{\cB}{{\cal B}}
\newcommand{\cH}{{\cal H}}

\newcommand{\1}{{\mathbf 1}}
\newcommand{\rc}{{\rm anti}}

\newcommand{\kN}{{\textsc N}}

\begin{document}

\title{THE NUMBER OF ORTHOGONAL CONJUGATIONS}
\author{Armin Uhlmann}
\date{University of Leipzig, Institute for
Theoretical Physics\footnote{As a pensioner}}

\maketitle

\begin{abstract}
After a short introduction to anti-linearity, bounds
for the number of orthogonal (skew) conjugations
are proved. They are saturated if the dimension of the
Hilbert space is a power of two. For other
dimensions this is an open problem.

keywords: Anti-linearity, canonical Hermitian form, (skew) conjugations.
\end{abstract}

\section{Introduction} \label{H0}
The use of anti-linear or, as mathematicians call it, of conjugate
linear operators in Physics goes back to E.~P.~Wigner, \cite{Wig32}.
Wigner also discovered the structure of anti-unitary operators,
\cite{Wi60}, in finite dimensional Hilbert spaces. The essential
differences to the linear case is the existence of 2-dimensional
irreducible subspaces so that the Hilbert space decomposes into a
direct sum of 1- and 2-dimensional invariant spaces in general.
Later on, F.~Herbut and M.~Vuji{\v c}i{\' c} could clarify the
structure of anti-linear normal operators, \cite{HV66}, by proving
that such a decomposition also exists for anti-linear normal
operators. While any linear operator allows for a Jordan
decomposition, I do not know a similar decomposition of an
arbitrary anti-linear operator.

In the main part of the paper there is no discussion of what is
happening in case of an infinite dimensional Hilbert space. There
are, however, several important issues both in Physics and in
Mathematics: A motivation of Wigner was in the prominent
application of (skew) conjugations, (see the next section for
definitions), to time reversal symmetry and related inexecutable
symmetries. It is impossible to give credit to the many
beautiful results in Elementary Particle Physics and in Minkowski
Quantum Field Theory in this domain. But it is perhaps worthwhile
to note the following: The CPT-operator, the combination of
particle conjugation C, parity operator P, and time-reversal T,
is an anti-unitary operator acting on bosons as a conjugation
and on fermions as a skew conjugation. It is a genuine symmetry of
any relativistic quantum field theory in Minkowski space. The
proof is a masterpiece of R.~Jost, \cite{RJ65}. There is a further
remarkable feature of anti-linearity shown by CPT. This operator
is defined up to the choice of the point $\mathbf{x}$ in Minkowski
space on which PT acts as $\mathbf{x} \to - \mathbf{x}$. Calling
this specific form CPT$_{\mathbf{x}}$, one quite forwardly
shows that the linear operator ${\rm CPT}_{\mathbf{x}} {\rm
CPT}{_\mathbf{y}}$ is representing the translation by the vector
$2(\mathbf{x} - \mathbf{y})$.

The particular feature in the example at hand is the splitting
of an executable symmetry operation into the product of two
anti-linear ones. This feature can be observed also in some
completely different situations. An example is the
possibility to write the output of quantum teleportation,
as introduced by Bennett et al \cite{BBCJPW93}, \cite{NC00, BZ06},
as the action
of the product of two anti-linear ones on the input state vector,
see \cite{Uh00a, Uh03, BNT12}.

These few sketched examples may hopefully convince the reader
that studying anti-linearity is quite reasonable --- though
the topic of the present paper is by far not so spectacular.

The next two sections provide a mini-introduction to
anti-linearity. In the last one it is proved that the number
of mutually orthogonal (skew) conjugations is maximal if the
dimension of the Hilbert space is a power of two. It is
conjectured that there are no other dimensions for which this
number reaches its natural upper bound.

\section{Anti- (or conjugate) linearity} \label{H1}
Let $\cH$ be a complex Hilbert space of dimension
$d < \infty$. Its scalar product is denoted by
$\langle \phi_b, \phi_a \rangle$ for all
$\phi_a, \phi_b \in \cH$. The scalar product is assumed linear
in $\phi_a$. This is the ``physical'' convention going
back to E.~Schr\"odinger. $\1$ is the identity operator.\\
{\em Definition 1:}
An operator $\vartheta$ acting on a complex linear space
is called {\em anti-linear} or, equivalently, {\em conjugate
linear} if it obeys the relation
\begin{equation} \label{rule1}
\vartheta (\, c_1 \phi_1 + c_2 \phi_2 \,) =
c_1^* \vartheta \phi_1 + c_2^* \vartheta \phi_2
, \quad c_j \in \mathbb{C}  \; .
\end{equation}

As is common use, $\cB(\cH)$ denotes the set (algebra) of all
linear operators from $\cH$ into itself. The set (linear space)
of all anti-linear operators is called $\cB(\cH)_{\rc}$.
Anti-linearity requires a special definition of the
Hermitian adjoint.\\
{\em Definition 2 (Wigner):}
The Hermitian adjoint, $\vartheta^{\dag}$, of
$\vartheta \in \cB(\cH)_{\rc}$ is defined by
\begin{equation} \label{rule2}
\langle \phi_1 , \vartheta^{\dag} \,  \phi_2 \rangle  = \langle
\phi_2 , \vartheta \, \phi_1 \rangle, \quad
\phi_1, \phi_2 \in \cH \; .
\end{equation}

A simple but important fact is seen by commuting $\vartheta$
and $A = c \1$.
One obtains $(c \vartheta)^{\dag} = c \vartheta^{\dag}$,
saying: \, $\vartheta \to \vartheta^{\dag}$ {\em
is a complex linear operation,}
\begin{equation} \label{rule4}
\bigl( \sum c_j \vartheta_j \bigr)^{\dag} =
\sum c_j \vartheta^{\dag}_j  \; .
\end{equation}
This is an essential difference to the linear case:
{\em Taking the Hermitian adjoint is a linear operation.}

A similar argument shows, that the eigenvalues of an anti-linear
$\vartheta$ form circles around zero. If there
is at least one eigenvalue and $d>1$, let $r$
be the radius of the largest such circle. The set of all
values $\langle \phi, \vartheta \phi \rangle$, $\phi$ running
through all unit vectors, is the disk with radius $r$. $\bullet$
See \cite{HJ91} for the more sophisticated real case.

We need some further definitions.\\
{\em Definition 3:}
An anti-linear operator $\vartheta$ is said to be
{\em Hermitian} or {\em self-adjoint}
 if $\vartheta^{\dag} = \vartheta$.
$\vartheta$ is said to be {\em skew Hermitian} or
{\em skew self-adjoint} if $\vartheta^{\dag} = - \vartheta$.
The linear space of all Hermitian (skew Hermitian) anti-linear
operators are denoted by
$$
\cB(\cH)_{\rc}^{+} \, \hbox{ respectively }
\, \cB(\cH)_{\rc}^{-} \; .
$$

Rank-one linear operators are as usually written
\begin{displaymath}
(|\phi' \rangle\langle \phi''|) \, \phi :=
\langle \phi'', \phi \rangle \, \phi' ,
\end{displaymath}
and we define similarly
\begin{equation} \label{rank1a}
(|\phi' \rangle\langle \phi''|)_{\rc} \, \phi :=
\langle \phi, \phi'' \rangle \, \phi' \; ,
\end{equation}
projecting any vector $\phi$ onto a multiple of $\phi'$.
Remark that we do not use
$\langle \phi''|$ decoupled from its other part.
We do not attach any meaning to $\langle \phi''|_{\rc}$ as a
standing alone expression\footnote{Though one could do so
as a conjugate linear form.} !

An anti-linear operator $\theta$ is called a unitary one
or, as Wigner used to say, an anti-unitary, if
$\theta^{\dag} = \theta^{-1}$. A {\em conjugation} is an
anti-unitary operator which is Hermitian, hence fulfilling
$\theta^2 = \1$. The anti-unitary $\theta$ will be called a
{\em skew conjugation} if it is skew Hermitian,
hence satisfying $\theta^2 = - \1$.

\section{The invariant Hermitian form} \label{H2}
While the trace of an anti-linear operator is undefined,
the product of two anti-linear operators is linear. The trace
\begin{equation} \label{Hform1}
(\vartheta_1 , \vartheta_2) := \T \, \vartheta_2 \vartheta_1
\end{equation}
will be called the {\em canonical Hermitian form,} or just the
{\em canonical form} on the the space of anti-linear operators.

An anti-linear $\vartheta$ can be written uniquely as a sum
$\vartheta = \vartheta^{+} + \vartheta^{-}$ of an Hermitian
and a skew Hermitian operator with
\begin{equation} \label{def3}
\vartheta \to \vartheta^{+} :=
\frac{\vartheta + \vartheta^{\dag}}{2}  , \quad
\vartheta \to \vartheta^{-} :=
\frac{\vartheta - \vartheta^{\dag}}{2} \; .
\end{equation}
Relying on (\ref{Hform1}) and (\ref{def3}) one concludes
\begin{equation} \label{rule5}
(\vartheta^{+} , \vartheta^{+}) \geq 0 , \quad
(\vartheta^{-} , \vartheta^{-}) \leq 0 , \quad
(\vartheta^{+} , \vartheta^{-}) = 0 .
\end{equation}

In particular, equipped with the canonical form,
$\cB(\cH)_{\rc}^{+}$ becomes an Hilbert space.
Completely analogue, $-(.,.)$ is a positive definite scalar
product on $\cB(\cH)_{\rc}^{-}$. Bases of these two Hilbert
spaces can be obtained as follows:

Let $\phi_1, \phi_2, \dots$ be a basis of $\cH$. Then
\begin{equation} \label{plusb}
(|\phi_j \rangle\langle \phi_j|)_{\rc} , \quad
\frac{1}{\sqrt 2} ((|\phi_j \rangle\langle \phi_k|)_{\rc}
 + (|\phi_k \rangle\langle \phi_j|)_{\rc} ) \; ,
\end{equation}
where $j.k =1, \dots, d$ and $k < j$, is a basis of
$\cB(\cH)_{\rc}^{+}$ with respect to the canonical form.
As a basis of $\cB(\cH)_{\rc}^{-}$ one can use the anti-linear
operators
\begin{equation} \label{minusb}
\frac{1}{\sqrt 2} ((|\phi_j \rangle\langle \phi_k|)_{\rc}
 - (|\phi_k \rangle\langle \phi_j|)_{\rc} ) \; .
\end{equation}
By counting basis lengths one gets
\begin{equation} \label{dimension}
\dim \cB(\cH)_{\rc}^{\pm} = \frac{d(d \pm 1)}{2} \: .
\end{equation}
It follows: The {\em signature of the canonical Hermitian form
is equal to $d = \dim \cH$.} Indeed,
\begin{equation} \label{signature}
\dim \cB(\cH)_{\rc}^{+} - \dim \cB(\cH)_{\rc}^{-}
= \dim \cH \; .
\end{equation}

\section{Orthogonal (skew) conjugations} \label{H3}
The anti-linear (skew) Hermitian operators are the elements
of the Hilbert spaces $\cB(\cH)_{\rc}^{+}$ and
$\cB(\cH)_{\rc}^{-}$. Their scalar products are restrictions
of the canonical form (up to a sign in the skew case).
Therefore to ask for the maximal number of mutually orthogonal
conjugation or skew conjugations, is a legitim question.

These two numbers depend on the dimension $d = \dim \cH$
of the Hilbert space only. Let us denote by $N^{+}(d)$
the maximal number of orthogonal conjugations and by
$N^{-}(d)$ the maximal number of skew conjugations.
By (\ref{dimension}) it is
\begin{equation} \label{nr}
N^{\pm}(d) \leq \frac{d(d \pm 1)}{2} \: .
\end{equation}
To get an estimation from below, one observes that the tensor
products of two conjugation and that of two skew conjugations
are conjugations. Therefore
\begin{equation} \label{nrc1}
N^{+}(d_1 d_2) \geq N^{+}(d_1) N^{+}(d_2) + N^{-}(d_1) N^{-}(d_2)
\end{equation}
and, similarly,
\begin{equation} \label{nrs1}
N^{-}(d_1 d_2) \geq N^{+}(d_1) N^{-}(d_2) + N^{-}(d_1) N^{+}(d_2)
\end{equation}
because the direct product of two orthogonal (skew) conjugations
is orthogonal. Now consider the case that equality holds in
(\ref{nr}) for $d_1$ and $d_2$. Then one gets the inequality
\begin{displaymath}
N^{+}(d_1 d_2) \geq \frac{d_1(d_1 + 1) d_2(d_2 +1) +
d_1(d_1 - 1) d_2(d_2 -1)}{4}
\end{displaymath}
and its right hand side yields $d(d+1)/2$ with $d = d_1 d_2$.
Hence there holds equality in (\ref{nrc1}). A similar
reasoning shows equality in (\ref{nrs1}) if equality holds
in (\ref{nr}). Hence: {\em The set of dimensions for which
equality takes place in (\ref{nr}) is closed under multiplication.}

To rephrase this we call ${\kN}_{\rc}$ the set of
dimensions for which equality holds in (\ref{nr}):

{\em If $d_1 \in {\kN}_{\rc}$ and $d_2 \in {\kN}_{\rc}$
then $d_1 d_2 \in {\kN}_{\rc}$}.\\
$2 \in {\kN}_{\rc}$ will be shown by explicit calculations below.
Hence {\em every power of two is contained in ${\kN}_{\rc}$}.

Let us shortly look at $\dim \cH = 1$.
It is $N^{+}(1) = 1$ and $N^{-}(1) = 0$. Indeed, any
anti-linear operator in $\mathbb{C}$ is of the form
$\vartheta_a z = a z^*$. This is a conjugation if $|a|=1$.
There are no skew conjugations. The canonical form reads
$(\vartheta_a, \vartheta_b) = a^* b$.

\noindent \underline{Conjecture:} ${\kN}_{\rc}$ consists of
the numbers $2^n$, $n=0,1,2, \dots$.

Skew Hermitian invertible operators exist in even dimensional
Hilbert spaces only. Therefore, no odd number except 1 is
contained in ${\kN}_{\rc}$. This, however, is a rather trivial
case. Already for $\dim \cH = 3$ the maximal number $N^{+}(3)$
of orthogonal conjugations seems not to be known.

\subsection{$\dim \cH = 2$}
To show that $2 \in {\kN}_{\rc}$ one chooses a basis
$\phi_1,\phi_2$ of the 2-dimensional Hilbert space $\cH$
and defines
\begin{eqnarray} \label{dim2a}
\tau_0 (c_1 \phi_1 + c_2 \phi_2) & = & c_1^*\phi_2 - c_2^* \phi_1
, \\
\tau_1 (c_1 \phi_1 + c_2 \phi_2) & = & - c_1^*\phi_1 + c_2^* \phi_2
 , \\
\tau_2 (c_1 \phi_1 + c_2 \phi_2) & = & i c_1^*\phi_1 + i c_2^* \phi_2
 , \\
\tau_3 (c_1 \phi_1 + c_2 \phi_2) & = & c_1^*\phi_2 + c_2^* \phi_1
\; .
\end{eqnarray}
For $j,k \in \{ 1,2 \}$ and $m \in \{ 1,2,3 \}$ one gets
\begin{displaymath}
\langle \phi_j, \tau_m \phi_k \rangle
= \langle \phi_k, \tau_m \phi_j \rangle
\end{displaymath}
saying that these anti-linear operators are Hermitian. One also
has $\tau_m^2 = \1$ for $m \in \{ 1,2,3 \}$. Altogether,
$\tau_1, \tau_2, \tau_3$ are conjugations. To see that they are
orthogonal one to another we  compute
\begin{eqnarray} \label{dim2b}
\tau_1 \tau_2 \tau_3 = - i \tau_0, & \quad & \tau_1 \tau_2 =
i \sigma_3 \; , \\
\tau_3 \tau_1 = i \sigma_2 ,  & \quad & \tau_2 \tau_3 = i \sigma_1
\; ,
\end{eqnarray}
and
\begin{equation} \label{dim2c}
\tau_2 \tau_0 = \sigma_2, \quad \tau_1 \tau_0 = \sigma_1,
\quad \tau_3 \tau_0 = \sigma_3
\end{equation}
The trace of any $\sigma_j$ is zero. Because of (\ref{dim2b})
and (\ref{dim2c}) we see, that $\tau_1, \tau_2, \tau_3$ is an
orthogonal set of conjugations while $\tau_0$ is a skew
conjugation. Now $N^{+}(2) = 3$ and $N^{-}(2) = 1$ as was
asserted above.

\underline{Remark}: There is a formal difference to the
published version \cite{hconju4}:
In the present version the sign has been changed in the
definition of $\tau_1$ to get the more symmetrical relations
(\ref{dim2b}) and (\ref{dim2c}).
$$ $$
\underline{acknowledgements}  Thanks to the referee and to
B.~Crell for helpful support.

 \end{document}